\documentclass{aa}
\usepackage{graphicx}
\usepackage{epstopdf}
\usepackage{natbib}
\usepackage{url}

\newcommand{\dotarsec}{.\hspace{-0.9mm}'\!\hskip0.4pt'\hspace{-0.2mm}}

\begin{document}
   \title{Discriminant analysis of solar bright points and faculae}

   \subtitle{II. Contrast and morphology analysis}

   \author{P. Kobel \inst{1,2} \and J. Hirzberger \inst{1} \and S. K. Solanki \inst{1,3} }

   \offprints{P. Kobel}

   \institute{Max-Planck Institut f\"{u}r Sonnensystemforschung, Max-Planck-Stra\ss e 2, 37191 Katlenburg-Lindau, Germany\\
              \email{philippe.kobel@a3.epfl.ch}
              \and Ecole Polytechnique F\'{e}d\'{e}rale de Lausanne, 1007 Lausanne, Switzerland \\
              \and School of Space Research, Kyung Hee University, Yongin, Gyeonggi, 446-701, Korea}

   \date{...}


  \abstract
      {Taken at a high spatial resolution of $\sim 0\dotarsec1$, Bright Points (BPs) are found to coexist
with faculae in images and the latter are often resolved as adjacent striations. Understanding the properties
of these different features is fundamental to carrying out proxy magnetometry. }
      {To shed light on the relationship
between BPs and faculae, we studied them separately after the application of a classification method, developed and described
in a previous paper) on active region images at various heliocentric angles.}
      {In this Paper, we explore different aspects
of the photometric properties of BPs and faculae, namely their G-band contrast profiles, their peak contrast in
G-band and continuum, as well as morphological parameters.}
      {We find that:
(1) the width of the contrast profiles of the classified BPs and faculae are consistent with studies of disk center BPs at
and limb faculae, which indirectly confirms the validity of our classification,
(2) the profiles of limb faculae are limbward skewed on average, while near disk center they exhibit both
centerward and limbward skewnesses due to the distribution of orientations of the faculae,
(3) the relation between the peak contrasts of BPs and faculae and their apparent area discloses a trend
reminiscent of magnetogram studies.
}
      {The skewness of facular profiles provides a novel constraint for 3D MHD models of faculae.
As suggested by the asymmetry and orientation of their contrast profiles, faculae near disk center could be induced by
inclined fields, while apparent BPs near the limb seem to be in fact small faculae misidentified.
The apparent area of BPs and faculae could be possibly exploited for proxy magnetometry.
}


   \maketitle
%

\section{Introduction}
\label{sec_intro}

Solar photospheric Brights Points (BPs) and faculae are known to be the bright radiative signatures of
small-scale kilo-Gauss (kG) magnetic flux concentrations near disk center and near the limb \citep{Mehltretter74,
Wilson81}, respectively. Since they can be observed in broad spectral bands with short exposure times ($\sim 10$
ms), so that the atmospheric aberrations are frozen in the image, BPs and faculae play an important role in
indirectly tracing such flux concentrations at high spatial and temporal resolution \citep{Schuessler_rev92}.
Yet the interpretation of these observations requires a deeper understanding of the phenomena responsible for the
appearance of BPs and faculae, if we aim at performing actual ``proxy magnetometry''.

Kilo-Gauss magnetic features are generally thought to be well described by magnetic flux tubes or flux sheets \citep[e.g.][]{Solanki93, Voegler05, Yelles09}. According to the basic ``hot-wall'' picture \citep{Spruit76}, BPs are seen when the line of sight (LOS) is
parallel to the flux tube axis, i.e. directed toward the depressed ``bottom'' of the optical depth unity surface
inside the tube, whereas faculae arise from an inclined view on the hot granular ``wall'' of that surface. Provided that most kG flux tubes are nearly vertical due to buoyancy \citep[][]{Schuessler88, Jafarzadeh14}, they are mostly viewed
from overhead at disk center and obliquely near the limb. This basic picture has been confirmed in its
salient points by 3D radiative MHD simulations at disk center
\citep[e.g.][]{Schuessler03, Shelyag04} and near the limb \citep[][]{Keller04, Carlsson04}. However, the
diagnostics of the resulting synthetic images reveal remaining quantitative discrepancies with the observations
regarding the peak contrast, size and ``striated appearance'' of faculae \citep{Keller04, Steiner07}.

On the observational side, the classical approach to test and constrain the ``hot-wall'' model of BPs and faculae
has been to study the Center-to-Limb Variation (CLV) of the contrast. According to that model, the contrast
varies from center to limb as the hot wall becomes ever more visible \citep[][]{Spruit76}, but the many
measurements of
``facular contrast'' CLVs (meant as the contrast of all photospheric brightenings, i.e. of BPs and faculae
together) give rather controversial results \citep[see][for a review]{Steiner07}. The presence
of different proportions of BPs and faculae in the datasets could possibly play a role. Except for the ``extreme
limb'' (heliocentric angles $> 60^\circ$), high-resolution images indeed disclose mixtures of BPs and faculae
\citep{Hirz05, Berger07}, with an increasing fraction of faculae toward the limb \citep[][hereafter referred to as Paper I]{Kobel09}.
The coexistence of BPs and faculae at the same disk position could be explained by the inclination of the fields
with respect to LOS, whereby BPs would be induced by flux tubes aligned along the LOS and faculae by inclined
flux tubes \citep[as proposed by][see also Paper I]{Keller04}.

Moreover, it is not clear either whether the BPs and faculae seen at different heliocentric angles are
manifestations of similar magnetic features (field strength, sizes). Early high-resolution observations raised
the suspicion that faculae (referred to as ``facular granules'') were the signatures of bigger magnetic
features than BPs \citep[see review of][]{Muller85}. Such a suspicion was supported by 2D MHD calculations which
showed that small flux tubes were mainly bright at the center of the disk \citep{Deinzer84}, while larger ones
brightened mainly near the limb \citep{Knoelker88b}. At a very high spatial resolution of around
$0\dotarsec1$, however, facular granules are found to break into smaller ``striations'', seemingly associated
with gaps of reduced field strength \citep{DePontieu06, Berger07}. This morphological change of
faculae at higher resolution raises new questions, and to our knowledge, there has been no systematic comparison of
the properties of these facular elements with those of BPs.


To clearly separate the observational properties of BPs and faculae, we apply the following approach on high-resolution images of active regions at several heliocentric
angles. First, previously-segmented bright ``magnetic'' features are assigned to be either BPs or faculae via an algorithm based on Linear Discriminant Analysis (see Paper I for details). Then, their
contrast and morphology is analyzed separately, while addressing the following questions:
\begin{itemize}
\item How do the contrast and morphology of BPs and faculae differ?
\item How do the photometric properties of these two classes of features vary from center to limb?
\end{itemize}

We first analyzed the information contained in G-band contrast
profiles of BPs and faculae in Sect. \ref{sec_profiles}. Then, we considered and compared the values of G-band
and continuum contrast of those features in Sect. \ref{sec_contrast}. Finally, Sect. \ref{sec_morphology}
presents a characterization of the morphology of individual BPs and faculae elements.
Although our photometric-based classification method is not yet fully calibrated (in the absence of joint magnetogram data), the statistical results do provide fresh information, allow a comparison with literature values and
provide new constraints on models and MHD simulations.


\section{Detection and discrimination of BPs and faculae}\label{sec_method}

\subsection{Dataset}\label{sec_dataset}

Our dataset consists of simultaneous G-band (430 $\pm$ 0.5 nm) and G-continuum (436 $\pm$ 0.5 nm) filtergrams of
active regions, recorded at the 1m--Swedish Solar Telescope (SST). They cover seven disk positions in the range
$0.56 \leq \left< \mu \right> \leq 0.97$ (see Table \ref{table_class}), where $\left< \mu \right>$ denotes the
center of each field of view (FOV). At each disk position, only one to three image pairs recorded at instants
of best seeing were selected, and reconstructed by phase-diversity. The resulting images reach almost
diffraction-limited quality (angular resolution $\sim 0\dotarsec1$, see Paper I for examples). For each image
pair, the brightness contrast was defined relative to the mean intensity $\left< I \right>_{\rm QS}$ of a
selected ``quasi-quiet'' area, cospatial in G-band and continuum, as $C = (I - \left< I \right>_{\rm QS})/ \left<
I \right>_{\rm QS}$. The G-band and continuum contrast are herein subscripted $C_{\rm G}$ and $C_{\rm C}$, respectively.

\subsection{BPs and faculae segmentation}\label{sec_segmentation}

The principles and details of the segmentation algorithm are presented in Paper I. In brief, a
Multi-Level-Tracking (MLT) segmentation \citep{Bovelet01} was first carried out on the G-band images, after the
application of a high-pass spatial filter. The MLT levels were finely spaced in order to resolve groups of BPs
and faculae striations into \emph{individual elements}. Then, we removed the segmented features that did not
contain a minimum of 5 pixels with $C_{\rm G}$ above a given threshold, $C_{\rm G,t}$, and $C_{\rm G}-C_{\rm C}$
above another threshold, $C_{\rm diff,t}$. The choice of these thresholds relies on the brightness excess of
bright ``magnetic'' features in G-band compared to continuum \citep{Berger98, Shelyag04}. Because this G-band
brightness excess decreases towards the limb while the absolute contrast increases, $C_{\rm G,t}$ was raised in a
systematic way for lower $\left< \mu \right>$ (see Paper I). In contrast, $C_{\rm diff,t}$ was kept constant for
all disk positions, as granulation shows only little G-band contrast excess regardless of $\left< \mu \right>$.

\subsection{Characteristic profiles and classification as BPs and Faculae}\label{sec_classification}

In order to classify the segmented features as BPs and faculae, we developed a method based on the \emph{Linear
Discriminant Analysis} \citep[LDA,][]{Fischer36}, using a selected reference sample of features (``training set'')
with chosen discriminant parameters (see Paper I for details). The training set consisted of 200 BPs and 200
faculae, selected by visual inspection at 5 different disk positions.

\begin{figure}
\centering
\includegraphics[width=\columnwidth]{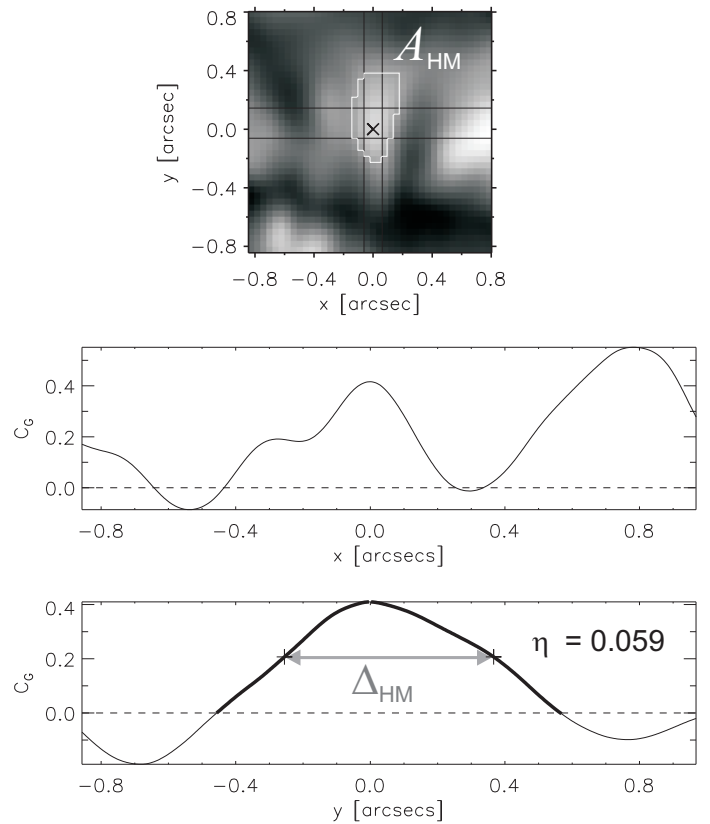}
\caption{Example of local frame orientation and G-band contrast profiles of a facular element located at $\mu \sim 0.9$.
Top panel: Orientation of the feature in its local $x/y$ coordinate frame. The black lines delimit the pixels having
$C_{\rm G} \geq 0.9 C_{\rm G,max}$ used for profile averaging. The pixels having $C_{\rm G} \geq 0.5 C_{\rm
G,max}$, with total area $A_{\rm HM}$ (see Sect. \ref{sec_contrastsize}), are contoured in white.
Lower panels: Average G-band profiles along $x$ and $y$. The
retrieved \emph{characteristic profile}, in this case chosen along y, is overplotted in thick. $\Delta_{\rm HM}$
represents its width at half-maximum ($C_{\rm G} = 0.5 C_{\rm G,max}$).
The value of the skewness $\eta$ of the characteristic profile is indicated (see Sect. \ref{sec_asym}).}
\label{fig_profile}
\end{figure}

To obtain discriminant parameters for the LDA, we retrieved G-band contrast profiles along directions specific to
each feature, such that these profiles adequately characterize BPs and faculae. To this end, each feature was oriented
in a local coordinate frame $x/y$, corresponding to the principal axes of its ``G-band contrast moment of
inertia'' and such that the latter is minimized about the y axis (cf. Paper I). Averaged G-band contrast profiles were then extracted along $x$ and $y$,
restricted to their positive contrast values, and the smoothest of these profiles was designated as the
\emph{characteristic profile}. We emphasize that these profiles were retrieved \emph{from the spatially-filtered
images}, to allow the definition of the same reference level ($C_{\rm G}=0$) for the characteristic profiles of
all features. Figure \ref{fig_profile} gives an example of a feature oriented in its $x/y$ frame, as well as the
retrieved averaged profiles in $x$ and $y$ and its characteristic profile. The width and the mean slope at the
edges of these characteristic profiles were found to be suitable discriminant parameters, together with the
feature area defined by segmentation (see Paper I for the definition and ``discriminant power'' of these
parameters).

LDA was then carried out in order to find a unique discriminant variable as a linear combination of the above
parameters, and such that it would best discriminate the training set classes according to Fischer's criterion
\citep{Fischer36}. By choosing a judicious threshold on that variable, all the segmented features at each disk
position of our dataset were ultimately classified as BPs or faculae. To lower the rate of misclassifications and
eliminate the features which are difficult to classify as BPs or faculae, the features
whose value of the discriminant variable were contained within a given range about the threshold were left
unclassified (this range was chosen such as to reject the same fraction equal to 0.2 of the BPs and faculae of the
training set).

\begin{figure*}
\centering
\includegraphics[width=\textwidth]{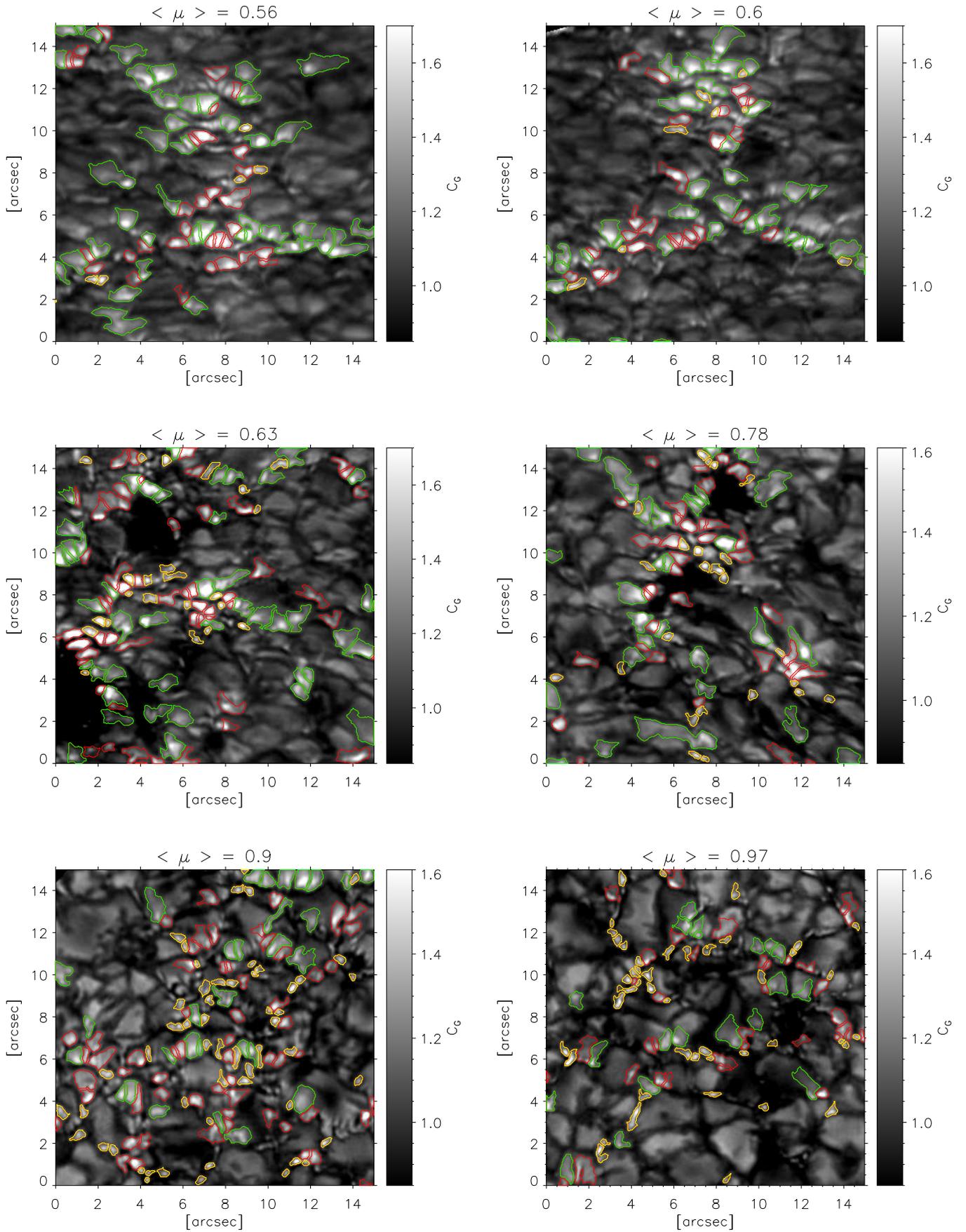}
\caption{Subfields extracted from the G-band images at various $\left<\mu\right>$, illustrating the CLV of the
appearance and relative proportion of the features classified. The faculae are contoured
in green, the BPs in yellow and the unclassified features in red. The contours correspond to the border of the
features as defined by the segmentation map (corresponding to the lowest MLT level $C_{\rm G} = 0$). The
direction of the closest limb is towards the top for all subfields.}
\label{fig_contours}
\end{figure*}

The results of this classification with rejection are summarized in Table \ref{table_class}. As a
complement, Fig. \ref{fig_contours} shows the contours of classified features in subfields extracted at various
$\left< \mu \right>$, thereby giving a visual impression of which features were classified as BPs and faculae,
and the variation of their appearance with $\left< \mu \right>$.
It should be stressed that our classification relied on purely photometric parameters, and therefore can only
sort features \emph{appearing} rather as BPs or as faculae.

\begin{table}
\caption{Classification results at each $\left< \mu \right>$ of our dataset, where $\left< \mu \right>$ refers to
the value at the center of the field of view of the images.
$N_{\rm tot}$ and $N_{\rm rej}$ are
the total number of segmented features and the number of rejected ones (unclassified). $N_{\rm bp}$ and $N_{\rm
fac}$ stand for the number of classified BPs and faculae, while ${N_{\rm bp}}^\dag$ and ${N_{\rm fac}}^\dag$ are
the numbers of BPs and faculae used for the contrast CLVs (see Sect. \ref{sec_contrastclv}).}
\vskip3mm
\label{table_class}
\centering
\begin{tabular}{c c c c c c c}     
\hline\hline
$\left< \mu \right>$ & $N_{\rm tot}$ & $N_{\rm rej}$ & $N_{\rm bp}$ & $N_{\rm fac}$ & ${N_{\rm bp}}^\dag$ & ${N_{\rm fac}}^\dag$\\
\hline
0.97 $\pm$ 0.003 & 1123 & 421 & 484 & 218 & 385 & 180 \\
0.94 $\pm$ 0.003 & 447 & 167 & 113 & 167 & 80 & 139 \\
0.9 $\pm$ 0.005 & 561 & 249 & 150 & 162 & 148 & 158 \\
0.78 $\pm$ 0.01 & 346 & 132 & 60 & 154 & 59 & 154 \\
0.63 $\pm$ 0.01 & 947 & 389 & 130 & 428 & 130 & 428 \\
0.6 $\pm$ 0.01 & 697 & 312 & 109 & 276 & 109 & 276 \\
0.56 $\pm$ 0.01 & 294 & 131 & 25 & 138 & 25 & 138 \\
\hline
all &  4415 & 1801 & 1071 & 1543 & 936 & 1473 \\
\end{tabular}
\end{table}

\section{Analysis of contrast profiles}
\label{sec_profiles}

To begin with, we investigated the orientation, width and shape of the G-band \emph{characteristic profiles} of
the classified BPs and faculae. As these profiles were retrieved along a direction yielding characteristic
profile shapes for BPs and faculae (see Sect. \ref{sec_classification}), the characteristic profiles are good candidates for comparison with radiative
transfer calculations accross 2D flux sheets \citep{Deinzer84, Knoelker88, Knoelker91, Steiner05}. In general,
the method for retrieving characteristic profiles yields profiles that are across the short dimension of elongated BPs, and along faculae striations.

\subsection{``Orientation'' of BPs and faculae}
\label{sec_orientation}

To understand the coming results related to such profiles, it is instructive to determine how the direction of
these characteristic profiles is distributed around the ``radial'' direction, i.e. the direction parallel to the
radius vector joining the disk center to the closest limb.
For this purpose, we defined $\phi \in (-90^\circ, 90^\circ)$ as the angle between the chosen direction of the characteristic profiles ($x$ or $y$ axis of its
oriented local frame) and the radial direction, as illustrated in Fig. \ref{fig_asymsign}.
The radial direction in our images is defined with an accuracy of 1$^\circ$ to 2$^\circ$.

\begin{figure*}
\centering
\includegraphics[width=\textwidth]{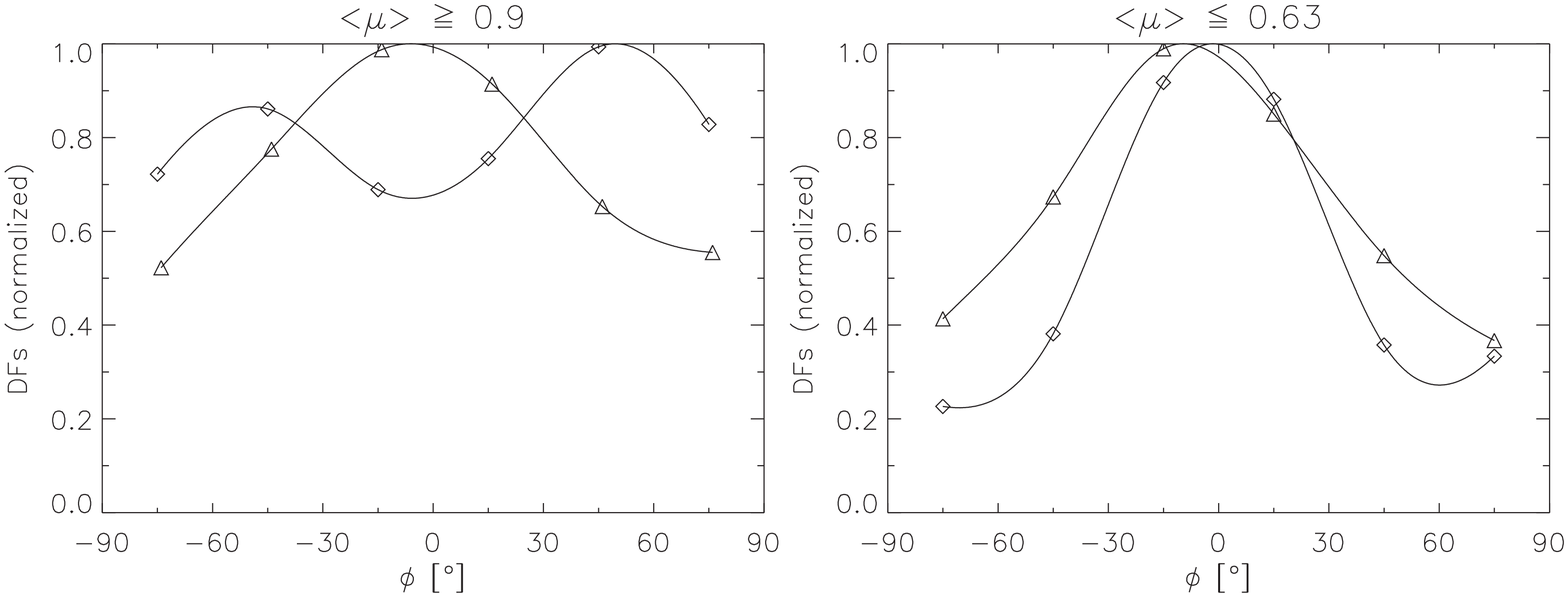}
\caption{Left: Density function (DF) of the angle $\phi$ between the chosen axis of the characteristic profile ($x$ or
$y$) and the radial direction for BPs ($\diamond$) and faculae ($\triangle$) in the ``center'' data ($\left< \mu
\right> \geq 0.9$). Right: Same for the ``limb'' data ($\left< \mu \right> \leq 0.63$). The symbols are plotted
centered in the bins. Cubic spline interpolations are overplotted as guides, and used to normalize the DFs to their maximum values. }
\label{fig_phi_DFs}
\end{figure*}

Because the orientation of BPs and faculae differs at disk center and near the limb, we separated the data into a
``center'' group (images with $0.9 \leq \left< \mu \right> \leq 0.97$, see Table \ref{table_class}) and a
``limb'' group ($0.56 \leq \left< \mu \right> \leq 0.63$). The normalized density functions (DFs) of the BPs and
faculae in these two groups are shown in Fig. \ref{fig_phi_DFs}.

The DF of faculae peaks close to $\phi = 0^\circ$ for the limb group, which was expected since facular brightenings are
on average radially oriented near the limb. Their DF
exhibits a similar but broader peak for the group near disk center. The preference for radial orientation of the profiles
near disk center is induced by the LOS inclination as well, since the ``center'' data actually cover a
range of heliocentric angles between 14$^\circ$ and 28$^\circ$. The broad wings of the distribution are
probably induced by inclined fields (cf. Sect. \ref{sec_intro}), whose orientation is unrelated to the radial
direction. At $\mu > 0.9$ a magnetic feature inclined by $20^\circ$ to the vertical \citep[e.g.][]{Jafarzadeh14, Martinez97} can produce a facula oriented in a significantly different direction than the radial one. A significantly lower $\mu$ such a typical inclination hardly affects the orientation.

The BPs near disk center have a wide distribution of orientation. The noticeable peaks at $\phi \sim \pm
50^\circ$ here are in fact not systematic; the shape of the DF varies from one image to the
next. Surprisingly though, the characteristic profiles of the BPs are dominantly radially oriented near the limb,
which strongly hints at possible misclassifications. A careful look at the classified BPs in the images near the
limb (see Fig. \ref{fig_contours}, $\left< \mu \right> \leq 0.63$) reveals that many of them appear to lie on the
edge or on a fragment of a granule. This suggests that these limbward BPs are in fact small faculae having a
BP-like appearance due to the lack of resolution. They could thus be misclassified by our purely photometric
segmentation algorithm (see the Discussion in Paper I).

\subsection{Width of BPs and faculae}
\label{sec_width}

Next, we investigated the width at half maximum of the characteristic profiles, $\Delta_{\rm HM}$
(illustrated in Fig. \ref{fig_profile}), in order to obtain a measure of the size of BPs and faculae and to
compare these with the literature.

\begin{figure*}
\centering
\includegraphics[width=\textwidth]{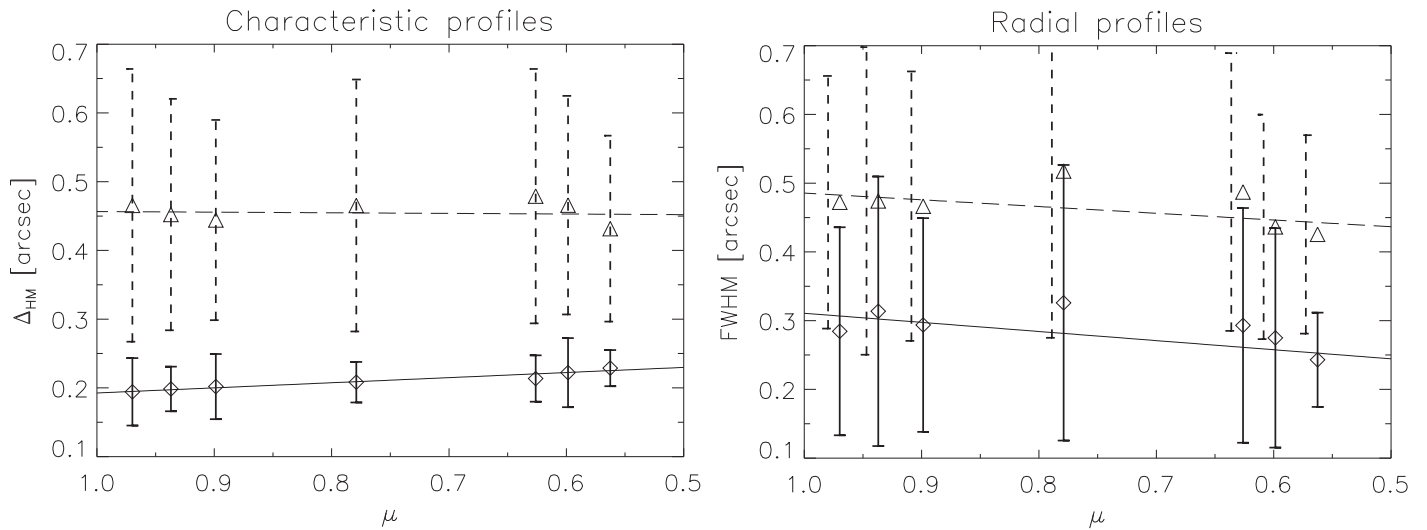}
\caption{Left: Center-to-limb variation (CLV) of the average $\Delta_{\rm HM}$ of BPs ($\diamond$) and faculae
($\triangle$), retrieved from the characteristic profiles.
Right: CLV of the average full-width at half maximum (FWHM) retrieved from radial profiles.
The error bars represent the standard deviations of BPs (solid) and faculae (dashed) at each $\left< \mu \right>$.
Linear regressions have been overplotted for BPs (solid) and faculae (dashed).}
\label{fig_hmw_CLV}
\end{figure*}

We first retrieved the average $\Delta_{\rm HM}$ at each $\left< \mu \right>$, in order to obtain a CLV for BPs
and faculae. As shown in Fig. \ref{fig_hmw_CLV} (left), the average $\Delta_{\rm HM}$ take clearly distinct
values for BPs and faculae at all $\left< \mu \right>$, the latter being roughly twice as wide. Also, these
averages are rather independent of $\left< \mu \right>$, in that any dependence lies within the dispersion. The difference between
the BP and faculae values as well as their $\mu$-invariance stems from both our classification method and from
the orientation of the characteristic profiles. The discriminant variable, on the basis of which the features are
classified, is indeed essentially based on the width of the characteristic profiles, and the classification
threshold on that variable was the same for all $\left< \mu \right>$ (see Paper I).
To illustrate the effect of the orientation of the characteristic profiles, Fig. \ref{fig_hmw_CLV} (right) shows
the analogous CLV of the full width at half maximum (FWHM, defined as well at $C_{\rm G} = 0.5 C_{\rm G,max}$)
obtained from \emph{radial} profiles. It can be seen that the $\Delta_{\rm HM}$ values of BPs and faculae are
more constant and have less dispersion than the FWHM of radial profiles, so that they represent a more robust
measurement of the size of the features. This is particularly true for BPs.

\begin{figure}
 \centering
 \includegraphics[width=\columnwidth]{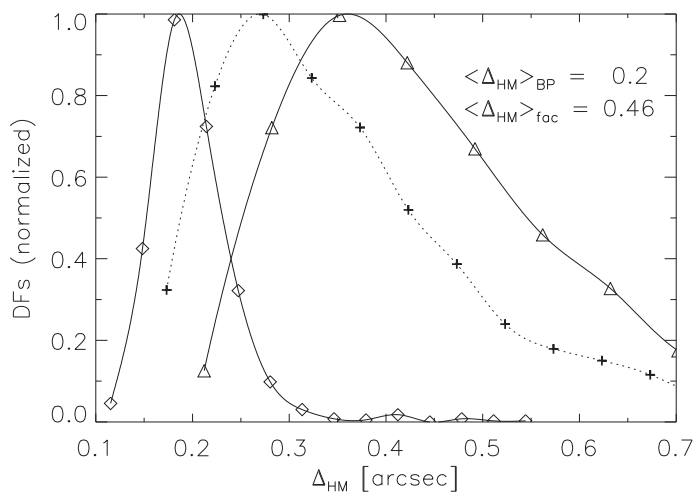}
  \caption{Normalized DFs of $\Delta_{\rm HM}$ for all BPs ($\diamond$) and faculae ($\triangle$). For comparison,
the normalized DF of all segmented features (BPs, faculae and rejected features) in the ``limb'' data is plotted as
well (+) with a dotted cubic spline.}
\label{fig_hmw_DF}
\end{figure}

The apparent $\mu$-invariance of $\Delta_{\rm HM}$ justifies plotting single DFs for BPs and faculae by combining
the features from all $\left< \mu \right>$ (Fig. \ref{fig_hmw_DF}). The histogram of BPs can be compared to the
one of \citet{Berger95}, who also retrieved FWHM of G-band profiles along feature-specific directions (in their case
chosen such as to minimize the width of BPs) for BPs in active regions at disk center. Whereas the mode of our BP
distribution corresponds to $\Delta_{\rm HM} = 0\dotarsec18$, their histogram peaks at around $0\dotarsec3$,
which is explained by the difference in  the spatial resolution of both studies \citep[the data of][were recorded at
the 50 cm Swedish Vacuum Solar Telescope, SVST]{Berger95}.
At a similar resolution as ours, \citet{Wiehr04} and \citet{Puschmann06} retrieved the distribution of BP diameter
by a MLT segmentation of G-band images from the SST. Despite their different definitions of ``size'', they obtain
modal values of $0\dotarsec 22$ and $0\dotarsec15$ respectively, similar to ours. Our mean value $\left< \Delta_{\rm HM} \right> = 0.2$ is also consistent with the mean BP diameter of $0\dotarsec 23$ derived from an analysis of G-band data from the Hinode Solar Optical Telescope \citep{Utz09}.
Finally, \citet{Crockett10} determined a log-normal distribution of the area of BPs using G-band images recorded at the 76 cm Dunn Solar Telescope. Even though the mode of their distribution corresponds to a larger BP diameter of $0\dotarsec31$, this can be partly explained by their choice of the brightest surrounding lane as threshold for the BP area, whereas we measured $\Delta_{\rm HM}$ at the half-maximum level of the profiles.

The $\Delta_{\rm HM}$ distribution of faculae peaks at larger values of about $0\dotarsec36$ with an average of
$\sim 0\dotarsec46$, which is consistent with the large extent of facular brightenings (up to $0\dotarsec5$ or
more) reported by \citet{Lites04}. However, our average value is larger than the average value of $0\dotarsec37$
obtained by \citet{Berger07}, using radial cuts at $\left< \mu \right> = 0.6$. We verified that this difference
is accounted for by the inclusion of only the features classified as faculae in our statistics (i.e. only about one quarter of all the
features at $\left< \mu \right> = 0.6$, see Table 1), whereas their measurements were performed on all their
detected features. When including all segmented features of the limb data in the statistics (BPs, faculae, and
rejected features), yielding the additional DF in Fig. \ref{fig_hmw_DF}, we obtain a similar average $\Delta_{\rm
HM}$ of $\sim 0\dotarsec37$.

The consistency of our distributions of $\Delta_{\rm HM}$, obtained for the \emph{classified} BPs and faculae at
\emph{all} the $\left< \mu \right>$ of our dataset, with other studies performed exclusively on BPs
(at disk center) or faculae (near the limb) indirectly supports the validity of our classification method.

\subsection{Asymmetry of facular profiles}
\label{sec_asym}

Of particular interest for the comparison with theoretical models are the asymmetries of
contrast profiles over the length of the faculae. These would pose new constraints that ``any model of faculae must
satisfy'' \citep{Steiner07}.
2D flux tube models indeed predict asymmetric facular intensity profiles, with a steep intensity rise induced by
the hot wall and a gentle fading on the limb side \citep{Deinzer84,Knoelker88,Knoelker91,Steiner05}.
Even though radial contrast profiles of limb faculae have been earlier retrieved \citep{Hirz05, Berger07}, there
has been no quantitative investigation of their asymmetry.

\begin{figure}
  \centering
  \includegraphics[width=0.75\columnwidth]{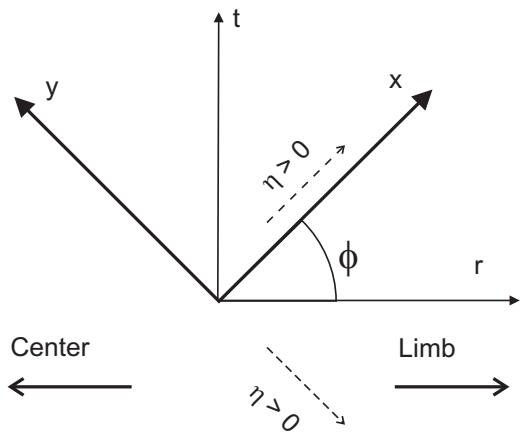}
  \caption{Illustration of the sign convention for $\eta$.
The $r$ and $t$ axes indicate the radial direction and the transverse one.
The $x$ and $y$ axes represent the local reference frame of a given feature.
Whether the characteristic profile of that feature is retrieved along $x$ or $y$, $\eta$ is counted positive if
the profile is skewed toward the limb (dashed arrows).
The angle $\phi$ is also represented, for the case in which the characteristic profile si along $x$.}
\label{fig_asymsign}
\end{figure}

To quantify the asymmetry of the characteristic profiles, we measured their \emph{skewness} as the third
standardized moment of these profiles:
\begin{eqnarray}
\eta = \frac{\int {\rm d}s \tilde{C}(s) (s - \left< s \right>)^3}{[\int {\rm d}s \tilde{C}(s) (s - \left< s \right>)^2]^{3/2}},\\
\tilde{C}(s) = \frac{C(s)}{\int \textrm{d}s C(s)}, \\
\left< s \right> = \int {\rm d}s \tilde{C}(s) s,
\end{eqnarray}
where $C(s)$ stands for the G-band characteristic profile and $s$ for the chosen coordinate of this profile
($x$ or $y$ axis of its local frame, cf. Sect. \ref{sec_classification}).
To compare with flux tube models predicting a limbward asymmetry, we counted the skewness $\eta$ as positive in
the direction of the characteristic profile pointing limbward (as illustrated in Fig. \ref{fig_asymsign}).

\begin{figure*}
\centering
\includegraphics[width=\textwidth]{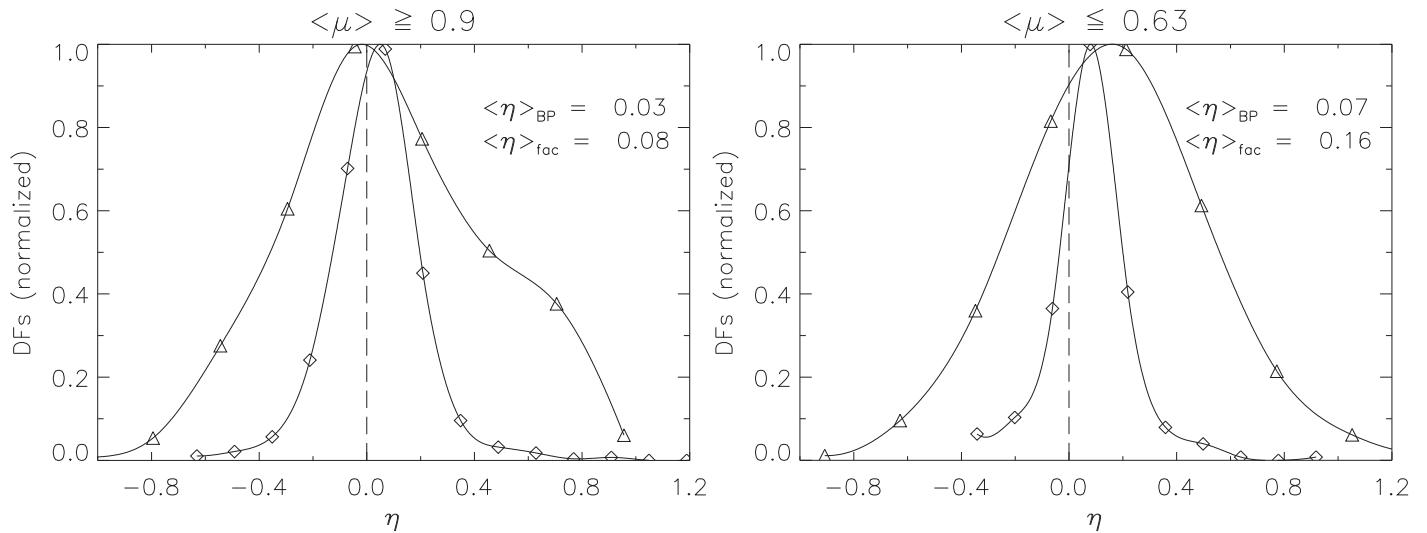}
\caption{Left: Normalized DFs of $\eta$ for BPs ($\diamond$) and faculae ($\triangle$) of the center data ($\left< \mu
\right> \geq 0.9$). Right: The same for the limb data ($\left< \mu \right> \leq 0.63$). A vertical dashed line ($\eta = 0$) separates positive and negative skewness.}
\label{fig_asym_DFs}
\end{figure*}

The ``center'' and ``limb'' data were treated separately, as for the study of the orientation. The DFs of the obtained
$\eta$ of faculae and BPs are displayed in Fig. \ref{fig_asym_DFs}. In the limb data, the faculae profiles are
clearly positively skewed (i.e. toward the limb) with an average value (and similar modal value) of 0.16.
In contrast, the BPs identified near the limb have an average value of 0.07 only. The fact that
their profiles are slightly limbward skewed as well further suggests that part of these BPs are misclassified
small faculae.

In contrast to the limb data, the faculae of the center data exhibit positive and negative values of $\eta$ in
similar proportions, with an average value of 0.08. This suggests that the sign of the asymmetry of facular
profiles is related to the orientation of their characteristic profiles, as the latter have a wider distribution of
orientation relative to the radial direction in the center data than in the limb data (see Fig.
\ref{fig_phi_DFs}). The slight excess of positive skewness can be explained by the rather wide range of $\mu$ of
the ``center'' images (see Table \ref{table_class}), so that the faculae are still preferably radially oriented
in these images (see Sect. \ref{sec_orientation}). As expected, the BPs exhibit a rather narrow distribution with
a low mean value of 0.03.

\begin{figure*}
\centering
\includegraphics[width=\textwidth]{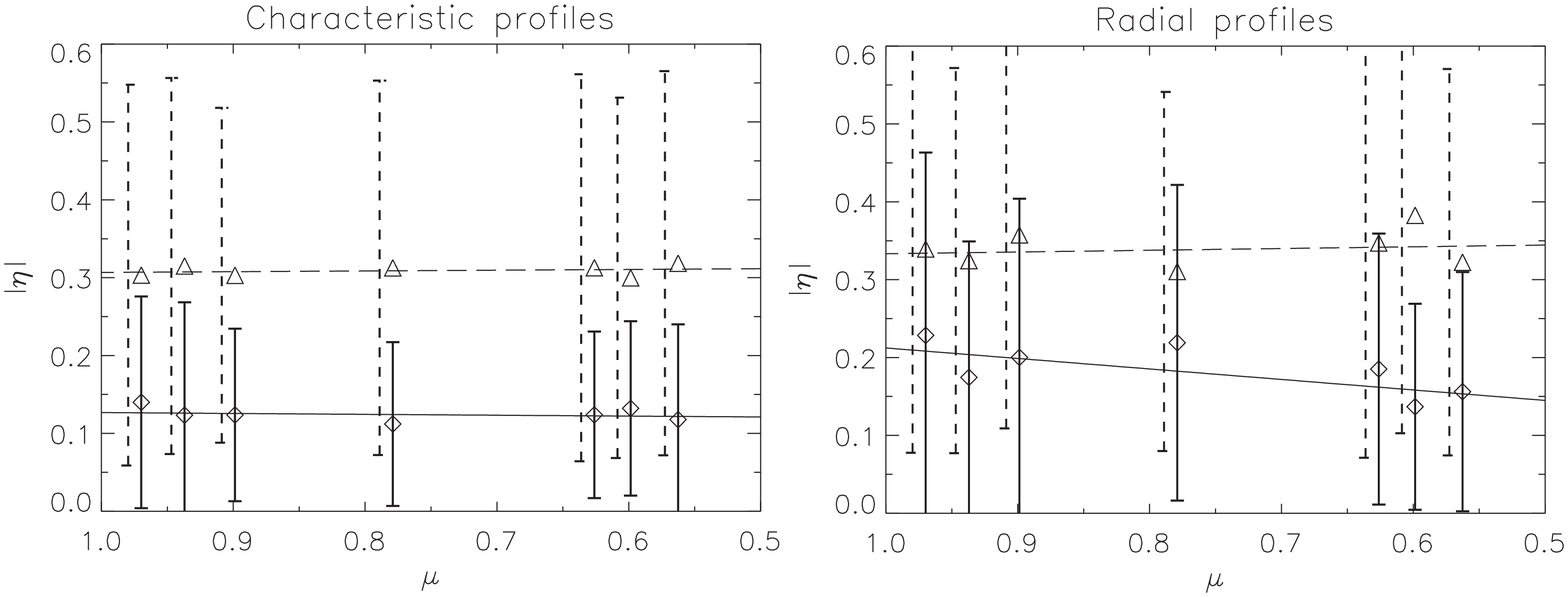}
\caption{CLV of the average absolute values of skewness $\eta$ of the characteristic profiles (left) and of radial profiles (right) for BPs ($\diamond$) and faculae ($\triangle$).
Error bars: Standard deviations of BPs (solid) and faculae (dashed). The
latter are left-shifted for clarity.
Solid and dashed linear regressions have been overplotted for faculae and BPs, respectively.}
\label{fig_asym_CLV}
\end{figure*}

Figure \ref{fig_asym_CLV} compares the CLV of the average $\mid\eta\mid$ measured on characteristic profiles and on radial profiles.
It is striking that both BPs and faculae reveal remarkably constant values of $\mid\eta\mid$ with $\left< \mu
\right>$ when measured on the characteristic profiles, whereas these values fluctuate much and have larger
standard deviations when measured on the radial profiles. The former are thus more robust observables to
use as constraint for models.

\section{G-band and continuum contrast}
\label{sec_contrast}

In this Section, we examine the behaviour of the G-band and continuum contrast of BPs and faculae by considering
the peak values of the features, $C_{\rm G,max}$, and $C_{\rm C,max}$ \footnote{In this Chapter, we used the
symbol $C_{\rm G,max}$ to refer both to the peak contrasts of the G-band characteristic profiles, measured from
the spatially filtered images, and the actual peak contrast of the features used in this Section, measured in the
non filtered images. The difference can be understood from the context and will be always clearly stated.} (taken
at the same pixel location and measured in the non-filtered images). Quantifying the contrast of the features via
their peak values rather than using averages avoids falsifying the facular contrast values by any granulation pixels mixed in the faculae.

\subsection{Center-to-limb variation}
\label{sec_contrastclv}

\begin{figure*}
\centering
\includegraphics[width=\textwidth]{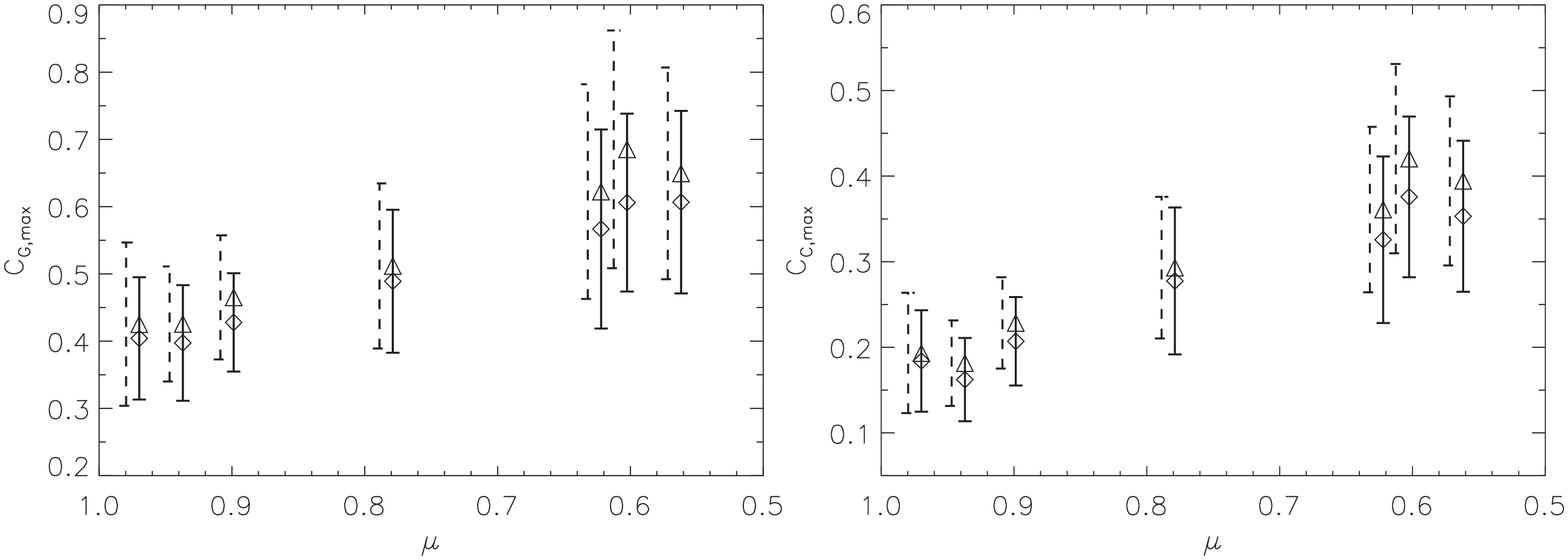}
\caption{Left: CLV of the averaged G-band peak contrast $C_{\rm G,max}$ of BPs ($\diamond$) and faculae
($\triangle$). Right: Idem in continuum, where $C_{\rm C,max}$ was taken at the same pixel location as $C_{\rm
G,max}$. To avoid bias (see main text), only features having $C_{\rm G,max} > 0.3$ were considered. The vertical
bars represent the standard deviations of BPs (solid) and faculae (dashed), the latter being artificially
left-shifted for readibility.}
\label{fig_Cclv}
\end{figure*}

We first looked for trends in the center-to-limb variation (CLV) of the peak contrast values of BPs and
faculae, averaged at each $\left< \mu \right>$ (Fig. \ref{fig_Cclv}).
To avoid a bias due to the progressive increase of the G-band threshold with
decreasing $\left< \mu \right>$ in our segmentation (see Sect. \ref{sec_method}), we imposed here an equal
threshold for all disk positions by requiring $C_{\rm G,max} > 0.3$ (corresponding to the highest value of the
$C_{\rm G,t}$ threshold used in the segmentation, see Paper I).

Whereas the peak contrast of faculae increases towards the limb as expected in our $\mu$ range (enhanced view on
the hot wall), the BPs surprisingly follow the trend of faculae, both in G-band and continuum. A straightforward
explanation for this common behaviour would be the misclassification of small faculae in the limb images ($\left<
\mu \right> \leq 0.63$). However, such misclassifications cannot account for the common contrast increase of BPs
and faculae at larger $\mu$ values ($\left< \mu \right> \geq 0.78$). Finally, note that faculae have
systematically larger contrast than BPs at all $\mu$ (both in G-band and continuum), especially in the limb
images, which can be due both to the hot-wall effect and the finite spatial resolution (faculae are on average larger, see Fig. \ref{fig_hmw_DF}, hence better resolved).

The similar behaviours of BPs and faculae implies that a distinction between these features would have a
minor influence in the contrast CLV of magnetic elements, at least in the considered $\mu$ range and
wavelengths.

The G-band peak contrast of BPs near disk center ($\left< \mu \right> = 0.97$) have an average of 0.42, similar to the value of 0.45 found by \citet{Zakharov05} with data taken at the SST as well. For comparison, BPs exhibit yet higher peak contrasts in the CN band at 388 nm, with average values of 0.6 obtained by \citet{Riethmueller10} with data from the SUNRISE observatory \citep{SUNRISE_mis}. Interestingly, their average BP continuum contrast at 525 nm reaches only 0.11, whereas our averaged G-continuum peak contast reaches 0.18.

\subsection{Size-dependence of the contrast}
\label{sec_contrastsize}

Because high-resolution imaging can provide ``direct'' measurements of the size of resolved magnetic features, it offers a way to directly probe the dependence of the contrast on the size of these
features. By comparison with flux tube models/simulations \citep[e.g.][]{Voegler05} and/or studies of contrast as a function of magnetogram
signal \citep[][]{Schnerr10, Kobel11, Kobel12, Yeo13}, such measurements could provide valuable constraints as well as deliver
a ``proof-of-principle'' of proxy magnetometry.

As a measure of the ``size'' of a feature, we considered the apparent area (projected onto the plane of the sky)
of the set of pixels whose G-band contrast in the spatially filtered images exceeds half of its maximum contrast,
$A_{\rm HM}$, illustrated in Fig. \ref{fig_profile}. Using the half-maximum level for
the measurement makes $A_{\rm HM}$ independent of the peak contrast of the features (as was the case for $\Delta_{\rm HM}$), and measuring $A_{\rm HM}$ in the
filtered images reduces the dispersion due to the medium and large-scale fluctuations of the intensity.

\begin{figure*}
\centering
\includegraphics[width=\textwidth]{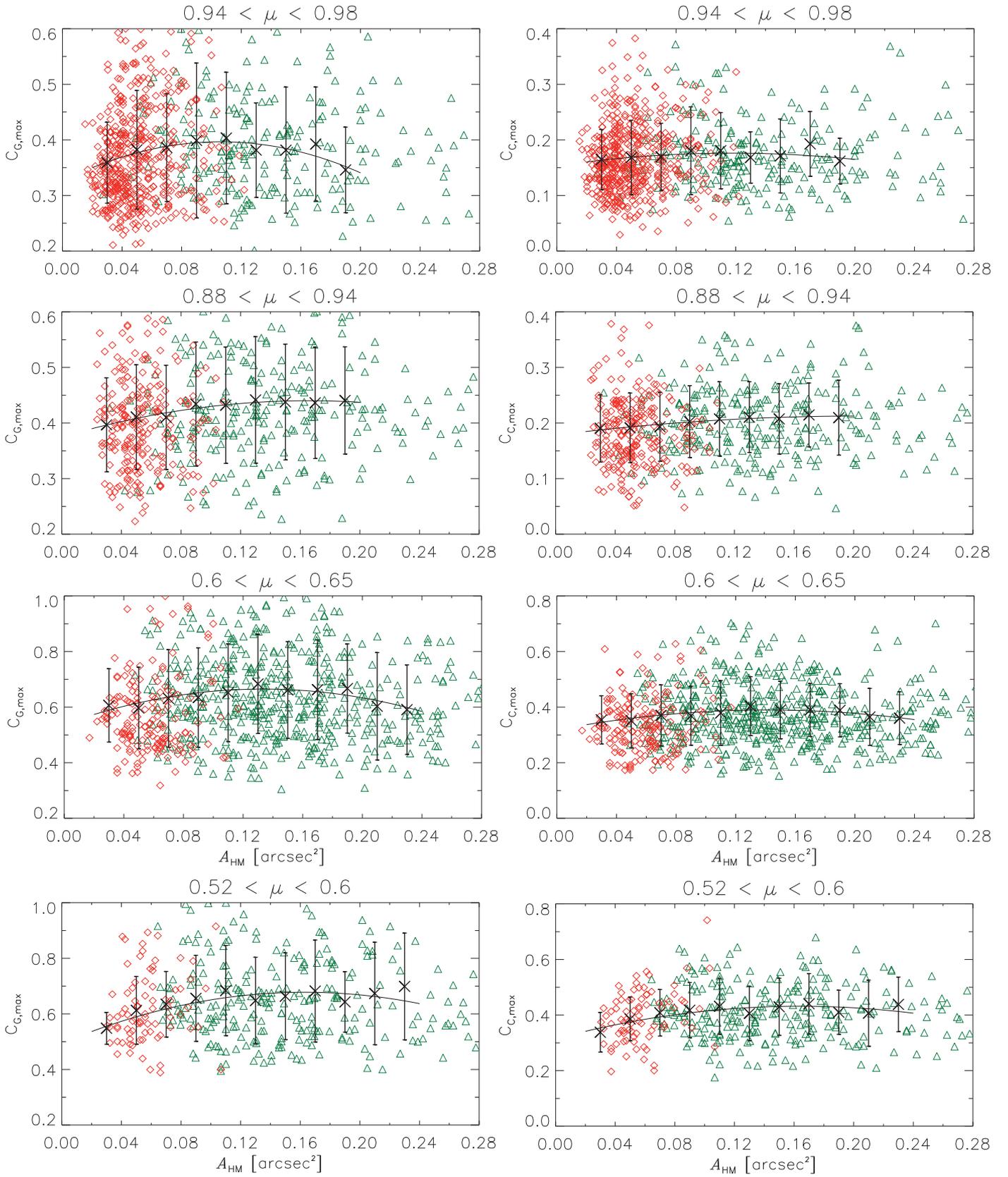}
\caption{Left column: Dependence of $C_{\rm G,max}$ on $A_{\rm HM}$ in four distinct intervals of $\mu$ near disk
center (upper two panels) and nearer the limb (bottom two panels). Right column: Same for $C_{\rm C,max}$.
Crosses and error bars: Average and standard deviation of the peak contrast of \emph{all} features (BPs, faculae
and rejected) in bins of $A_{\rm HM}$ of width 0.02 arcsec$^2$, restricted to bins having more than 10 contrast
points. Solid curve: Least-square quadratic fit of the average values (weighted by the inverses of the standard deviations).
The individual data points corresponding to BPs ($\diamond$, in red) and faculae ($\triangle$, in green) have been overplotted.}
\label{fig_GC_vs_Ahm}
\end{figure*}

Because of the CLV of the contrast and the geometrical foreshortening affecting $A_{\rm HM}$ with decreasing
$\mu$, we have divided the data into four intervals of $\mu$. Fig. \ref{fig_GC_vs_Ahm}
shows the scatterplots of $C_{\rm G,max}$ and $C_{\rm C,max}$ vs. $A_{\rm HM}$ for the BPs and faculae in the four
$\mu$ intervals.
In each interval, the contrast values of \emph{all} the features (BPs, faculae and rejected) were averaged in
bins of $A_{\rm HM}$ in order to identify an overall trend. This is justified by the relatively distinct ranges
of $A_{\rm HM}$ of BPs and faculae (owing to our classification), which allow a single trend to represent the
contrast behaviour of both types of features. Since we are not interested in comparing the magnitude of the contrast between different $\mu$ intervals
(unlike in Sect. \ref{sec_contrastclv}), we included here all the data points without imposing $C_{\rm G,max} >
0.3$.


Despite our segmentation of bright features only (see Sect. \ref{sec_segmentation}), the obtained trends bear a
striking resemblance with the analogous trends of contrast as a function of ``magnetogram signal'' (calibrated
measure of the Stokes $V$ amplitude), both for the continuum \citep[see][despite the differences in
wavelength]{Topka92, Lawrence93, Ortiz02, Kobel11, Kobel12} and the G-band \citep[see][]{Berger07}.
Thus, the concave shape of the trends as a function of $A_{\rm HM}$ becomes ever
more pronounced as $\mu$ decreases, in agreement with earlier studies. In addition, the present trends peak at larger $A_{\rm HM}$ for lower $\mu$
intervals, in the same way as the peak shifts to larger magnetogram signals.
To our knowledge, this qualitative analogy at various $\mu$ has not been unveiled before.

\begin{figure*}
\centering
\includegraphics[width=\textwidth]{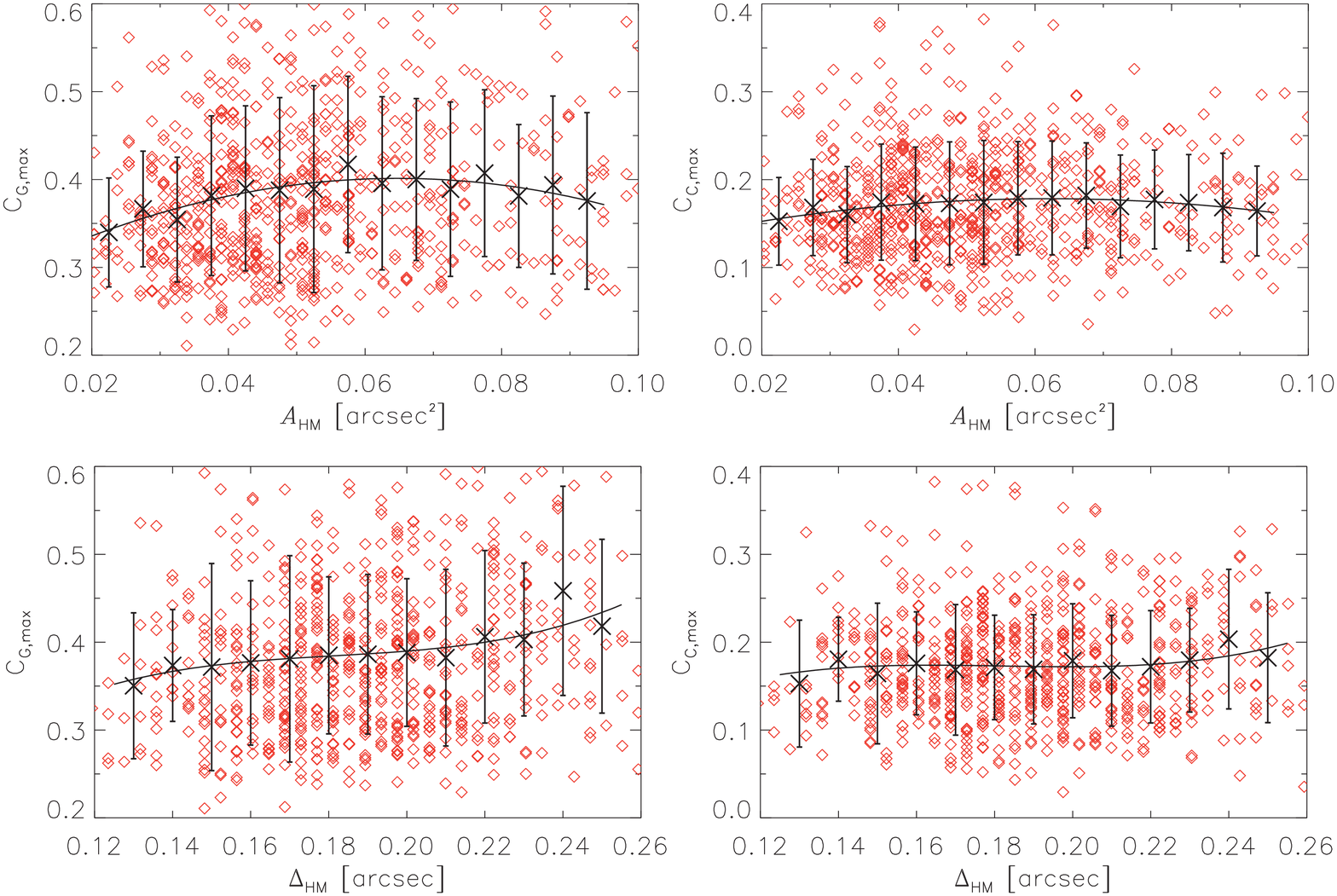}
\caption{Dependence of $C_{\rm G,max}$ (left) and $C_{\rm C,max}$ (right) on $A_{\rm HM}$ (upper panels) and
$\Delta_{\rm HM}$ (lower panels) for BPs in the ``center'' images ($\left< \mu \right> \geq 0.9$).
Crosses and error bars: Average and standard deviation of the BP peak contrasts in bins of $A_{\rm HM}$ of
0.005 arcsec$^2$ width and bins of $\Delta_{\rm HM}$ of 0.01 arcsec width.
Quadratic polynomials and cubic polynomials have been fitted to the average contrasts in bins of $A_{\rm HM}$ and
$\Delta_{\rm HM}$, respectively.
Individual contrasts of BPs are overplotted ($\diamond$).}
\label{fig_GC_vs_hm_BPs}
\end{figure*}

We now take a closer look at the BPs in the center data ($\left< \mu \right> \geq 0.9$). Since
BPs near disk center are thought to be the signatures of flux concentrations seen from overhead, it is expected
that their contrast--size relation closely follows the predictions of flux tube models. Such models predict a
decrease of the contrast with the size of the flux tubes (assuming a constant field strength), mainly because of
the increasing ratio between their internal volume and the heating surface of their ``walls''
\citep{Spruit76,Deinzer84,Fabiani92, Pizzo93}.
To verify this prediction, we presented in Fig. \ref{fig_GC_vs_hm_BPs} the dependence of the $C_{\rm G,max}$ and $C_{\rm C,max}$ of the
BPs on both $A_{\rm HM}$ and $\Delta_{\rm HM}$. The latter can be
considered a good proxy of the underlying flux tube/sheet width, since it is generally measured along the short
dimension of the BPs (owing to the orientation of the characteristic profiles).

The behaviour of the BP peak contrasts partly deviates from the theoretical expectations, both as a function of
$A_{\rm HM}$ and $\Delta_{\rm HM}$.
Except for the notable decrease of $C_{\rm G,max}$ and $C_{\rm C,max}$ as a function of $A_{\rm HM}$ for the large
BPs ($A_{\rm HM} > 0.06$ arcsec$^2$), the contrast exhibits an increase with $A_{\rm HM}$ for the smaller
BPs, in agreement with \citet{Viticchie10}. This ``anomaly'' could be due to the spatial resolution, affecting the $C_{\rm G,max}$ and $C_{\rm C,max}$
values of small BPs. \citet{Criscuoli09} demonstrated this effect by degrading the contrast of BPs computed from 2D flux sheet simulations. 
Note that \citet{Viticchie10} find that the contrast of BPs saturates when reaching a size for which they are properly resolved (for areas larger than 0.07 arcsec$^2$), whereas we observe a slight decrease of contrast for larger BPs (see also Fig. \ref{fig_GC_vs_Ahm}). This is probably due to their use of quiet Sun data only, as the investigation of contrast vs. magnetic flux by \citet{Kobel12} reveals a contrast saturation in the quiet Sun but a concave trend in active regions, even when considering only bright pixels.
That the initial increase of BP contrast with $A_{\rm HM}$ is due to spatial resolution also finds support in the resemblance with studies of contrast vs. magnetogram signal. Since the magnetograms have a much poorer spatial
resolution, the initial increase is reasonably explained by the increasing
``filling factor'' (fractional area of the resolution element) of unresolved features. However, true physical
effects cannot be entirely discarded, in particular the possible dependence of the field strength on the
size of the flux concentrations (see Sect. \ref{sec_discussion2}).
The trends as a function of $\Delta_{\rm HM}$ are especially intriguing, in that the contrast monotonously rises
over the whole range of size (with a slight cubic shape). It is probable that the decrease of contrast toward
larger cross--sectional areas of the BPs is ``hidden'' when viewing the data as a function of $\Delta_{\rm HM}$,
because the latter is limited to the width of intergranular
lanes while the area is free to vary in the perpendicular dimension.
Finally, we note that irrespective of being examined as a function of $A_{\rm HM}$ or $\Delta_{\rm HM}$, the
shape of the contrast trends seem to be amplified in G-band with respect to continuum. As a function of $A_{\rm
HM}$, the difference between continuum and G-band is not unlike the difference between continuum and line core
contrast as a function of magnetogram signal observed by \citet{Frazier71} (at 525 nm) and \citet{Title92} (at
676.8 nm), and is consistent with the G-band high-resolution measurements of \citet{Berger07}.

\section{Morphology}
\label{sec_morphology}

\begin{figure}
 \centering
\includegraphics[width=0.75\columnwidth]{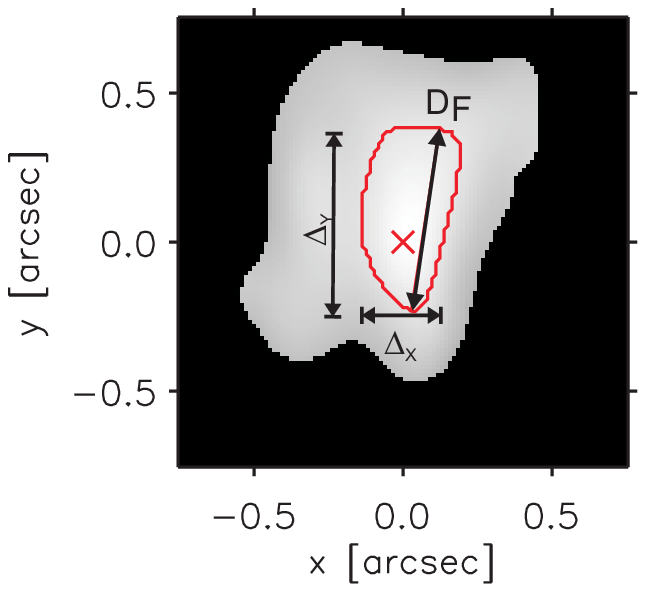}
  \caption{Illustration of a 2D figure used for the morphology studies. The feature is in its local $x/y$ reference
frame, and its contrast spatial distribution was bilinearly-interpolated by a factor three in each direction. The
boundary of the feature is defined by the segmentation map, with additional contour-smoothing. The inner contour
outlines the relevant figure composed by the pixels satisfying $C_{\rm G} > 0.5 C_{\rm G,max}$, and the oblique chord
corresponds to the ``Ferret'' diameter $D_{\rm F}$.}
\label{fig_figure}
\end{figure}

To study the morphological properties of BPs and faculae, we considered the 2D ``figures'' formed by the set of
pixels having G-band contrast values above half of the local maximum of the segmented features (in filtered
images), i.e. with $C_{\rm G} > 0.5 C_{\rm G,max}$ \footnote{Following \citet{Stoyan94}, these figures were
treated as simple two-dimensional objects, neglecting the fact that their pixels have different contrast
values.}. These figures are simply connected by virtue of the MLT segmentation, and their outline is by
definition independent of absolute contrast values. To minimize the spurious effect of pixellation (inducing
artificial roughness of the outline, erroneous areas and perimeters), the segmented features were
bilinearly-interpolated by a factor three in each direction with an additional contour smoothing, prior to the
extraction of the figures. The outline of such a figure is shown in Fig. \ref{fig_figure}, superposed on the
corresponding bilinearly interpolated feature.

\begin{figure*}
\centering
\includegraphics[width=\textwidth]{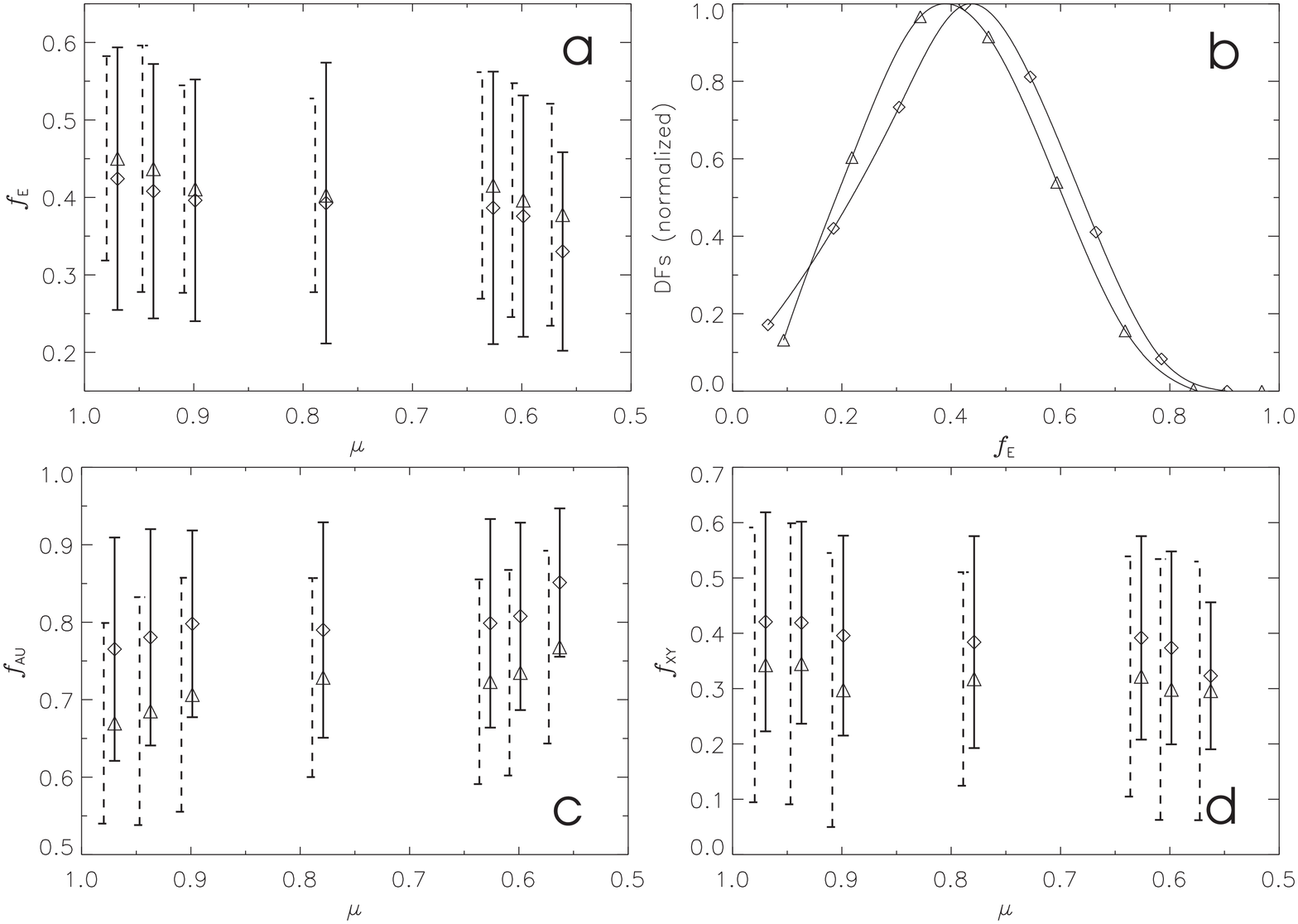}
\caption{a) CLV of the average ellipticities $f_{\rm E}$ of BPs ($\diamond$) and faculae ($\triangle$). b)
Normalized DFs of $f_{\rm E}$ for the BPs of the ``center'' data and the faculae of the ``limb'' data.
c) CLV of the average area-perimeter ratio $f_{\rm AU}$. d) The same for the width ratio $f_{\rm XY}$ in
the local reference frame $x/y$.
For all CLVs, the standard deviations of faculae (dashed) have been left-shifted for clarity.}
\label{fig_morphstat}
\end{figure*}

In the framework of geometrical set theory, the form of figures is adequately described by a variety of ``form
parameters'', defined as ratios independent of the position, scaling and orientation \citep{Stoyan94}. To characterize the ellipticity of each feature, we considered the ellipse whose
semi-major axis $a$ is given by the maximal chord length $D_{\rm F}$ of its associated figure, the so-called
``Ferret diameter'', and whose semi-minor axis $b$ is deduced from the figure area $A$ \footnote{Due to the
interpolation of the feature, the area $A$ of the figures considered here is not exactly equivalent to the area $A_{\rm
HM}$ considered in Sect. \ref{sec_contrastsize}.}:
\begin{equation}\label{eqn_ab}
a = \frac{D_F}{2}; b = \frac{A}{\pi a}
\end{equation}
The ellipticity factor can then be defined as:
\begin{equation}\label{eqn_ell}
f_{\rm E} = 1-\frac{b}{a} \in (0,1)
\end{equation}
%
This definition is indeed independent of the orientation and barely influenced by the irregularity
(non-convexity) of the figure contour. Fig. \ref{fig_morphstat}a displays the mean values of $f_{\rm E}$ for BPs
and faculae at each $\left< \mu \right>$. The values are roughly constant with $\left< \mu \right>$, although a
slight decrease towards the limb can be noticed for both BPs and faculae (this may be a possible foreshortening or spatial
resolution effect). Also, whereas the average values of BPs and faculae are rather similar, the former have
larger standard deviations, implying a larger variety of shapes.

For comparison with other studies, we computed the DFs of the ``center'' BPs and ``limb'' faculae, as shown in
Fig. \ref{fig_morphstat}b.
The faculae histogram can be directly compared with the one of \citet{Bovelet01}, who used an identical
definition of ellipticity, applied to MLT-segmented facular grains in a 658 nm continuum image at $\left< \mu
\right> = 0.54$ recorded at the SVST. Although not specified in their paper, the authors used as well a
normalized threshold at 0.5 for their figures (E. Wiehr, private communication). Despite their lower statistics (only 638
faculae extracted at only one $\left< \mu \right>$) and their lower spatial resolution (which is of little influence on
large structures as faculae, however), their histogram is very similar to ours. The mode of their distribution also corresponds to
$f_{\rm E} \sim 0.4$ and the distribution spans values from $f_{\rm E} = 0$ to 0.8. The ellipticities of disk
center BPs in active regions have been investigated previously by \citet{Berger95} on G-band images from SVST.
These authors used a slightly different definition, as the semi-axes ratio $a/b$ of a best-fitting ellipse to the
BP shape given by their segmentation map. They thus obtained a distribution with a mean value of 1.5 and a severe
drop for ellipticities larger than 2. If we transform our previous definition of $f_{\rm E}$ into $a/b$, we then
obtain a larger mean value of 1.88 as well as a more extended distribution. This discrepancy with \citet{Berger95} is most
probably related to the lower spatial resolution of the SVST, causing BPs to appear more circular.

Another useful shape ratio to quantify the deviations from circularity is the ``area-perimeter ratio'' defined as:
\begin{equation}\label{eqn_au}
f_{\rm AU} = \frac{4 \pi A}{U^2} \in (0,1)
\end{equation}
where $U$ stands for the perimeter of the figure, and $f_{\rm AU} < 1$ for all figures other than a perfect disc. Note that this definition is very sensitive to
small-scale deviations of convexity in the figure outline, through the second power of $U$ in Eq, \ref{eqn_au}. The larger average
values of BPs compared to faculae at each $\left< \mu \right>$ (see Fig. \ref{fig_morphstat}c) indicate that BPs
have more convex and regular shapes.
This is probably a consequence of the finite spatial resolution, as BPs are much smaller. In addition, the figure
contours of the faculae can be more complex, due to the possible
clustering of fine striations, which are not always properly resolved by the MLT segmentation.
Also, the finite resolution is probably responsible for the common increase of the average $f_{\rm AU}$ of BPs
and faculae toward lower $\mu$.

The two shape ratios presented above are strict ``form parameters'', in that they are independent of the feature's
orientation. However, this neglects the distribution of contrast within
the features, which contains information preferred directions such as the principal axes of the contrast momentt of inertia (see Sect.
\ref{sec_classification}). As all our BPs and faculae have already been oriented in a local $x/y$ frame according to
their ``contrast moment of inertia'', we can define an new shape ratio in that frame:
\begin{equation}\label{eqn_xy}
f_{\rm XY} = 1-\frac{\Delta_{\rm x}}{\Delta_{\rm y}}
\end{equation}
where $\Delta_{\rm x}$ and $\Delta_{\rm y}$ are the widths of the figure along $x$ and $y$, measured through the
location of maximum contrast $C_{\rm G,max}$ (see Fig. \ref{fig_figure}). Note that  $f_{\rm XY}$ increases with the elongation in $y$ with respect to $x$, as the features are preferably elongated along $y$ (since the $y$ axis is the one about which the contrast moment of inertia is minimum). Unlike for $f_{\rm E}$, BPs exhibit
slightly larger values of $f_{\rm XY}$ than faculae (Fig.
\ref{fig_morphstat}d). This difference can be attributed to the narrow contrast distribution of elongated BPs
about their $y$ axis, while faculae have a flatter, more diffuse distribution (this is also the case for adjacent
faculae striations due to the relatively weak brightness depressions between them). 
Moreover, the $f_{\rm XY}$ values of faculae display a larger dispersion than the BP ones, whereas it was the
opposite for $f_{\rm E}$. These differences can be attributed to
the proper orientation of BPs (reducing their dispersion) and to the more complex shape of the faculae (as
revealed by $f_{\rm AU}$), causing discrepancies between the axes of the ellipse used in the definition of
$f_{\rm E}$ (\ref{eqn_ab}) and $\Delta_{\rm x},\Delta_{\rm y}$. Hence, whereas the global ellipticity of BPs and
faculae is very similar when considered independently of their orientation, BPs appear slightly more elongated
when using a shape ratio that takes into account their contrast distribution.

\section{Discussion}
\label{sec_discussion}

\subsection{On the relationship between BPs and faculae}
\label{sec_discussion1}

Here we discuss some aspects of the differences and similarities between the features classified as BPs and
faculae, as revealed by our analysis.

We found that near the limb ($0.56 < \left< \mu\right> < 0.63$), faculae are preferably radially oriented (i.e. along the line joining the disk center to the nearest limb) and
their profiles exhibit a limbward skewness ($\eta > 0$).
A limbward asymmetry was predicted by 2D calculations \citep[][]{Deinzer84,
Knoelker88, Knoelker91} as a signature of the hot-wall effect.
Earlier observational studies of facular profiles nevertheless gave inconsistent results, although they did not statistically
quantify the asymmetry of the profiles.
Indeed, even though some isolated cases of asymmetric facular profiles were observed \citep{Lites04, Hirz05}, a
recent study found that the profile resulting from the average of the ``radial cuts'' of all detected bright
features at $\mu \sim 0.6$ was rather symmetric \citep{Berger07}. However, these authors did not distinguish BPs and faculae, and it is very likely that averaging all profiles smears out the asymmetry of
individual facular profiles.
Interestingly, the three radial profiles of faculae computed from the 3D MHD simulations of \citet{Keller04} do
not exhibit a visible asymmetry either. It should be borne in mind that the observed profiles are
\emph{projected onto the plane of the sky}, as the ones of \citet{Keller04}, whereas the profiles computed from
2D models are always shown as a function of the horizontal coordinate.
However, our facular profiles exhibit an asymmetric shape \emph{despite their projection onto the
plane of the sky}, and thus pose a novel constraint for the 3D simulations.

Near disk center, we found faculae to have a wider distribution of orientation (relative to the radial direction)
and to exhibit both limbward and centerward skewness. But interestingly, the ``degree of asymmetry'' of their
profiles (quantified through $\mid\eta\mid$) is very similar to that of the limb faculae.
This suggests that faculae near disk center are also produced by a ``hot wall effect'', but induced by the
inclination of the fields \citep[as suggested by][see also in Paper I]{Keller04}. Therefore the orientation of
their brightenings appears unrelated to the radial direction (although our ``center'' images cover a wide range
$0.9 < \left< \mu\right> < 0.97$ and contain as well LOS effects).

As expected, BPs near disk center have a large spectrum of orientation with very symmetric characteristic profiles.
Near the limb, surprisingly, the orientation of their characteristic profiles is preferably radial while
their profiles exhibit a slight limbward skewness.
This suggests that the features appearing as BPs near the limb are manifestations of the same hot wall effect as
limb faculae. But unlike large faculae, they take a BP-like appearance because they do not benefit from a large
granule on their limbward side, and due to the finite resolution. The appearance of faculae was indeed shown by
\citet{Keller04} to depend on the presence of a ``well-formed'' granule next to the flux concentrations. Such
very small facular brightenings were already reported by \citet{Lites04}.

The BP-like appearance of small faculae is also understandable in the light of our morphology study.
The statistics of the shape ratios indeed revealed that at the highest achievable
resolution, at which adjacent BPs and faculae striations can be resolved by MLT segmentation,
these elements bear only minor morphological differences (in spite of their large difference of size on average).
This similarity between BPs and faculae, in particular the relatively low
ellipticity of BPs, appears surprising when keeping in mind the often very complex structure of intergranular BPs in
active regions \citep[see][]{Berger04}, with the magnetic features behaving like a
``magnetic fluid'' in MHD simulation snapshots \citep{Schuessler03, Shelyag04}. However, a close look at the
subfields shown in Fig. \ref{fig_contours} (for $\left< \mu \right> = 0.9, 0.97$) reveals that many of these
complex structures are broken up into smaller adjacent ones by the MLT segmentation, while others are rejected
from the classification.
Moreover, even if the definitions of the shape ratios are scale-independent, their values depend very much
on the spatial resolution, especially for BPs.
A further increase in spatial resolution is thus expected to reveal more morphological differences.

The peak contrast of BPs and faculae was found to have a similar CLV. This is surprising since small and large flux tubes are predicted to have different CLVs of their contrasts \citep[e.g.][]{Knoelker88b}. A possible
explanation is that in the $\mu$ range considered here, the variation of contrast is not dominated by the
visibility of the hot wall, but by a ``limb darkening'' effect common to BPs and faculae. This could be due to
the geometrical shift of the optical depth combined with the larger temperature gradient outside than inside the
flux concentrations, causing a similar darkening of the quiet environment of BPs and faculae.

\subsection{Possible prospects for proxy magnetometry}
\label{sec_discussion2}

The trends foud here of the peak contrasts (in G-band and continuum) as a function of the area of the features in
different intervals of $\mu$ strikingly resemble the trends of the variation of the contrast as a function of
magnetogram signal \citep{Frazier71,Topka92,Lawrence93,Ortiz02, Kobel11, Kobel12, Yeo13}.

In magnetogram sudies, the spatial resolution is often poorer, so that the magnetic features are often
unresolved \citep[even for high resolution data, see][]{Riethmueller10} and their magnetogram signal (which scales with their magnetic flux) can
be roughly interpreted as a measure of their cross--sectional area (assuming a roughly constant field
strength). The contrast is then seen to increase with small magnetogram signals as the features become larger and
better resolved, until the contrast decreases for larger signals associated with darker features (e.g. micropores).

Note that studies of the contrast as a function of magnetogram signal average the contrast of unrelated pixels in
the images having magnetogram signals within given bins, and thus bear no information about the identity of
the magnetic features.
The obvious similarity between our results and magnetogram studies thus implies that (1) there must be a correlation
between the observed area of the segmented features and the associated magnetic flux, and (2) in a statistical
sense, pixel by pixel contrast studies provide similar information as studies investigating the contrasts of individual magnetic features.


The possible correlation between the projected area of bright features and the flux of the associated magnetic
elements is partly expected at disk center. Using G-band images and joint spectropolarimetric data at disk center, \citet{Viticchie10} found the BP area and the associated magnetic flux to be correlated. Near the limb, however, one would expect the facular brightness to be
rather a simple ``cumulative function of their projected area'' \citep{Berger07}. There could nevertheless be an intuitive relation between
the visibility of the bright granular walls of facular features and the associated magnetic flux. Namely, the
larger the flux concentrations the more visible are the walls (and the larger the projected area) but the cooler
is the atmosphere traversed by the LOS, which could contribute to reduce the contrast of larger features. The
relation between the extent of the facular brightness and the amount of flux is also supported by the cuts across
magnetograms and faculae of \citet{Berger07}, showing that even at $\mu \sim 0.6$, both quantities are roughly
co-spatial.

Hence, through their similarity with magnetogram studies, our results seem to validate the use of proxy
magnetometry at high resolution.
However, a true assessment of the relation between the measure of area used in our study and the magnetic flux of
the features requires a dataset with joint magnetograms, and this area measurement should be properly calibrated for
its use as a proxy.

Note that the finite spatial resolution plays an important role in our measurements, as even when restricting
the case to BPs near disk center, the theoretical decrease of contrast with area is only recovered for large BP (see Fig. \ref{fig_GC_vs_hm_BPs}).
This is not in contradiction with the fact that most BPs considered here have dimensions largely exceeding the
diffraction limit, as the point spread function (PSF) could significantly smear the contrast at the core of the
BPs.
In addition, observational indications exist that the ``true'' (i.e. corrected for the effects of resolution) contrast-size relation of BPs matches the theoretical
expectations \citep[see][]{Spruit81, Ortiz02}.
Physical effects could also come into play to explain that an increase of contrast with size (instead of the
expected decrease) is found in the low size range.
It could be that the features having small sizes also harbour weaker fields (therefore less opacity depression, less heat input through the side walls and hence lower contrast). That small concentrations
of flux can have weaker field strength could be accounted
for by some inhibition of their ``convective intensification'' \citep{Parker78, Grossman98} through lateral
heating \citep[as theoretically predicted by][]{Venka86, Rajaguru00}. Evidence for weaker field strength of small magnetic features is provided by \citet{Solanki96, Khomenko03, Orozco07c}.

\section{Summary}
\label{sec_summary}

In the present study we followed the classification approach developed in Paper I to sort BPs and faculae in
high-resolution images of active regions at various heliocentric angles, and statistically analyzed their
photometric properties (contrast and morphology), as well as their variation with heliocentric distance.
Here we briefly summarize the obtained results:
\begin{enumerate}

\item The statistical distribution of the width of BPs and faculae classified with our method is consistent with
previous studies performed on disk center-BPs \citep{Berger95,Wiehr04,Puschmann06,Bovelet08}, or on limb-faculae
only \citep{Lites04,Berger07}, which indirectly supports our classification.

\item Limb faculae ($0.52< \mu < 0.65$) are mostly radially oriented and their contrast profiles are
skewed limbward (both in mode and average), as predicted by 2D MHD calculations
\citep{Deinzer84,Knoelker88,Knoelker91,Steiner05}. Nearer the disk center, their orientation is more widely
distributed and the profiles exhibit both limbward and centerward skewness in similar proportions, but with
absolute values comparable to the ones near the limb. This indicates that faculae near disk center are induced by
a similar hot wall effect through inclined fields, and that the orientation of the facular brightenings is related to the azimuthal direction of
their inclination. The skewness of facular profiles provides a novel constraint for 3D MHD simulations of faculae.

\item BPs observed near the limb are most probably very small faculae having a BP-like appearance due to our limited spatial resolution and the
rather small granule on their limbward side.

\item In the $\mu$-range considered here, the G-band and continuum contrast of BPs and faculae increases
similarly from disk center to limb.

\item At the current spatial resolution of roughly 100 km, the
morphological parameters investigated here bear only minor differences between BPs and facular elements. At all $\mu$, BPs tend to be systematically more regular in shape and more elongated when measuring the elongation in the principal axes of the contrast moment of inertia. 

\item The relation between the peak contrasts and the apparent (projected) areas of BPs and faculae has a trend which is qualitatively similar to the studies of contrast as a function of magnetogram signal, both for the continuum \citep[][see as well Chapter
4]{Ortiz02, Topka92, Lawrence93} and the G-band \citep{Berger07}.
This similarity implies a tight relation between the areas of the features observed in high resolution images and
the flux of the associated magnetic elements, and thus opens an interesting avenue for proxy magnetometry.
\end{enumerate}

\begin{acknowledgements}
We are grateful to M. Sch{\"u}ssler for his fruitful comments on the relation between the G-band and continuum contrast.
\end{acknowledgements}

\bibliographystyle{aa}
\bibliography{biblio}

\end{document}